\documentclass[twocolumn,showpacs,amsmath,amssymb,superscriptaddress,floatfix]{revtex4-2}
\pdfoutput=1

\usepackage{amsmath}
\usepackage{amsfonts}
\usepackage{float}
\usepackage{ graphicx } 
\usepackage{ bm } 
\usepackage{ color }
\usepackage{epstopdf}

\newcommand{\bea}{\begin{eqnarray}}
\newcommand{\eea}{\end{eqnarray}}

\newcommand{\na}{n_{j,A}}
\newcommand{\nb}{n_{j,B}}
\newcommand{\nc}{n_{j+1,A}}

\newcommand{\ca}{c_{j,A}}
\newcommand{\cb}{c_{j,B}}
\newcommand{\cc}{c_{j+1,A}}

\newcommand{\cad}{c_{j,A}^{\dagger}}
\newcommand{\cbd}{c_{j,B}^{\dagger}}
\newcommand{\ccd}{c_{j+1,A}^{\dagger}}

\newcommand{\braket}[2]{\left\langle #1 \middle | #2 \right\rangle} 
\newcommand{\mele}[3]{\left\langle #1 \middle | #2 \middle | #3 \right\rangle} 
\newcommand{\ket}[1]{\left | #1 \right\rangle}

\begin{document}
\title{Correlations and dynamical quantum phase transitions in an interacting topological insulator}

\author{ Wing Chi Yu }
\email{wingcyu@cityu.edu.hk}
\affiliation{Department of Physics, City University of Hong Kong, Kowloon, Hong Kong}

\author{ P. D. Sacramento }
\email{pdss@cefema.tecnico.ulisboa.pt}
\affiliation{ \textit CeFEMA,
	Instituto Superior T\'ecnico, Universidade de Lisboa, Av. Rovisco Pais, 1049-001 Lisboa, Portugal}
\affiliation{\textit Beijing Computational Science Research Center, Beijing 100193, China}

\author{ Yan Chao Li }
\affiliation{ \textit College of Materials Science and
	Opto-Electronic Technology, University of Chinese Academy of
	Sciences, Beijing 100049, China }
\affiliation{\textit Beijing Computational Science Research Center, Beijing 100193, China}

\author{ Hai-Qing Lin }
\affiliation{\textit Beijing Computational Science Research Center, Beijing 100193, China}

\date{\today}

\begin{abstract}
Dynamical quantum phase transitions (DQPTs), which refer to the criticality in time of a quantum many-body system, have attracted much theoretical and experimental research interest recently. Despite DQPTs are defined and signalled by the non-analyticities in the Loschmidt rate, its interrelation with various correlation measures such as the equilibrium order parameters of the system remains unclear. In this work, by considering the quench dynamics in an interacting topological model, we find that the equilibrium order parameters of the model in general exhibit signatures around the DQPT, in the short time regime. The first extrema of the equilibrium order parameters are connected to the first Loschmidt rate peak. By studying the unequal-time two-point correlation, we also find that the correlation between the nearest neighbors decays while that with neighbors further away builds up as time grows in the non-interacting case, and upon the addition of repulsive intra-cell interactions. On the other hand, the inter-cell interaction tends to suppress the two-site correlations. These findings could provide us insights into the characteristic of the system around DQPTs, and pave the way to a better understanding of the dynamics in non-equilibrium quantum many-body systems.
\end{abstract}

\maketitle

\section{Introduction}
In recent decades, the study of non-equilibrium phenomena in quantum many-body systems has attracted much attention in multidisciplinary physics. On one hand, many non-equilibrium phases of matter such as many-body localization \cite{Anderson1958,Gornyi2005,Basko2006,Nandkishore2015,Alet2017,Abanin2017,Luitz2017} and discrete-time crystals \cite{Wilczek2012,Sacha2017,Sacha2015,Khemani2016,Else2016,Else2017,Huang2018,Yao2017,Yu2019,Estarellas2020} have been theoretically proposed and are believed to have potential applications in quantum computing. On the other hand, the advance in quantum simulators make realizing interacting quantum many-body models possible in real experiments. The dynamics of the systems can now be investigated in a controlled manner using platforms such as cold atoms, optical lattices, and superconducting qubits  \cite{Lloyd1996,Langen2015,Gross2017,Schreiber2015,Smith2016,Bordia2016,Choi2016,Zhang2017}. Despite the rapid developments in recent years, the understanding of non-equilibrium quantum many-body physics is still far from complete. Understanding non-equilibrium critically, namely the dynamical quantum phase transitions (DQPTs), is believed to play one of the key roles to unveil the underlying mechanism in non-equilibrium quantum many-body phenomena.

Dynamical quantum phase transitions were originally introduced in two different notions. The first type of DQPT refers to the behavior of the order parameter in the long-time regime \cite{Yuzbashyan2006,Sciolla2010}. Another type of DQPT refers to the non-analyticity in the transient time Loschmidt rate \cite{Heyl2013,Heyl2014}, which is defined as 
\bea
\lambda(t)=-\lim_{N\rightarrow\infty}\frac{1}{N}\ln |l(t)|^2,
\eea
where $N$ is the system size and $l(t)=\braket{\Psi(t)}{\Psi(0)}$ is the Loschmidt amplitude that measures the overlap between the initial state $\ket{\Psi(0)}$ and the time evolved state $\ket{\Psi(t)}$. The Loschmidt rate is analogous to the free energy density in statistical physics \cite{Heyl2013}. Recently, it was shown that these two types of DQPT are actually related in the transverse-field Ising model with long-range interactions \cite{Zunkovic2018}. In this work, we will focus on the later definition of DQPT. This is also a more easily accessible scenario in experiment as it only involves observing the system's dynamic in the transient real time regime. Since its introduction, DQPT has attracted much attention and has been theoretically predicted \cite{Heyl2018,Zvyagin2016,Vajna2014,Hagymasi2019,Lacki2019,Schmitt2015,Zache2019,Lahiri2019,Sedlmayr2018,Yang2019,Kosior2018,Kyaw2018} and experimentally observed \cite{Jurcevic2018,Flaschner2018,Tian2019,Guo2019} in a number of condensed matter systems.

Despite the Loschmidt rate provides us an effective indicator for the DQPT, the information about the time-evolved many-body state around the transition that can be drawn from it is still limited. On the other hand, there were studies using the order parameters of the underlying equilibrium quantum phase transitions to track the nature of the quantum phase in DQPT \cite{Heyl2013,Hagymasi2019,Homrighausen2017}. However, to what extent the local order parameters will be useful in a DQPT remains generally unclear. Proposals of dynamical topological order parameter (DTOP) are also introduced in attempt to characterize the topological properties of the quenched system's dynamics. For example, there is DTOP defined in terms of the Pancharatnam geometrical phase and it has been shown in a number of models that the DTOP is quantized and exhibits a $\pi$ jump at the DQPT \cite{Flaschner2018,Budich2016}. However, calculating this quantity in interacting systems may be complicated as it is defined in terms of the Loschmidt amplitude in the momentum space. Recently, another DTOP in terms of the time ordered Green's function has been proposed and its applicability has been demonstrated in signalling the DQPT in a high-energy physics model in the presence of interactions \cite{Zache2019}. The primary motivation of the present work is to investigate the interrelation of various correlation measures of the system with the Loschmidt rate as a measure of DQPTs.

In this work, the spinless Su-Schrieffer-Heeger (SSH) model, which is the simplest model that possesses a topological phase in equilibrium, and its real-time dynamics under a sudden quench is considered. In our previous works, we derived a quasi-local order parameter of the model in equilibrium \cite{Yu2019a}. Here, we investigate the time dependence of this quasi-local order parameter and its relation to the DQPT. The real-space time ordered Green's function is also analysed. We find that the minima of these correlation measures are connected to the Loschmidt rate and provides an alternative way to understand the nature of the non-equilibrium quantum states around the transition. The paper is organized as follows. In Section \ref{sec:model}, a brief introduction of the model and the order parameters in the equilibrium model is given. The quench dynamics and the correlation measures in the non-interacting case are analysed in Section \ref{sec:non-interacting}. In Section \ref{sec:interacting} and Section \ref{sec:interacting2}, we extend the study to the SSH model with inter-cell and intra-cell fermion-fermion interactions. A conclusion is presented in Section \ref{sec:conclusion}.

\section{The model and the order parameters for the system in equilibrium} 
\label{sec:model}


The Su-Schrieffer-Heeger model describes a dimerized chain of spinless fermion with anisotropic inter-cell and intra-cell hopping \cite{Su1979}. The Hamiltonian is given by 
\bea
\label{eq:H_non}
H_0 &=& -\mu\sum_j(\na+\nb)\\\nonumber
&&-w\sum_j\left[(1+\eta)\cbd\ca+(1+\eta)\cad\cb\right.\\ \nonumber
&&\hspace{10pt}\left.+(1-\eta)\ccd\cb+(1-\eta)\cbd\cc\right],
\eea
where $c^{\dagger}_{j,\alpha}$ ($c_{j,\alpha}$) creates (annihilates) a spinless fermion of type $\alpha=A,B$ at the $j$-th unit cell, and $n_{j,\alpha}=c^{\dagger}_{j,\alpha}c_{j,\alpha}$. The parameter $\mu$ is the chemical potential, $w$ is the hopping amplitude, $\eta$ characterizes the anisotropy in the inter-cell and intra-cell hopping. Unless otherwise specified, we consider the half-filling case where $\mu=0$ and take $w=1$.

The Hamiltonian in Eq. (\ref{eq:H_non}) can be diagonalized at $\eta=-1$ and $\eta=1$ respectively by transforming the spinless fermion operators into Majorana fermion operators. For $\eta>0$, all the Majorana fermions are paired up and the system is in a trivial state. On the other hand, for $\eta<0$, two Majorana fermions are decoupled at each of the edges. These combine to form fermionic edge modes and the system is in a topological phase. The system undergoes a topological quantum phase transition at $\eta=0$ \cite{Yu2016}.

Using the scheme proposed by Gu et. al. \cite{Gu2013} with some fundamental extensions, we derived the order operators that characterize the topological phase and the trivial phase in the model by analysing the reduced density matrix spectrum and the mutual information. The order operators read \cite{Yu2019a}
\bea
\label{eq:O-}
O_- &=& \frac{1}{2}(\ccd\cb+\cbd\cc)\\\nonumber
&&-\nb\nc+\frac{1}{2}(\nb+\nc),
\eea
and
\bea
\label{eq:O+}
O_+ &=& \frac{1}{2}(\cbd\ca+\cad\cb)\\\nonumber
&&-\na\nb+\frac{1}{2}(\na+\nb), 
\eea
where '$-$' and '$+$' denotes the topological and trivial phase respectively. The order operator for the topological phase is quasi-local in the sense that it involves sublattice from adjacent unit cells. In fact, it closely resembles the order parameter of a bond-order wave.

In the case with interactions, the Hamiltonian is given by $H=H_0+H_I$, where
\bea
H_I = U\sum_j\na\nb+V\sum_j\nb\nc,
\label{eq:HI}
\eea
and $U$ and $V$ characterises the inter-cell and intra-cell interaction strength respectively. The topological phase, as captured by the order parameter in Eq. (\ref{eq:O-}), is found to be robust to repulsive inter-cell interactions. It can also emerge in the case of positive $\eta$ in the presence of appropriate interactions. In addition, two types of charge-density waves (CDW) and phase separation (PS) are also observed in some parameter regimes \cite{Yu2016}.

%
%

\begin{figure}
	\includegraphics[width=0.5\textwidth]{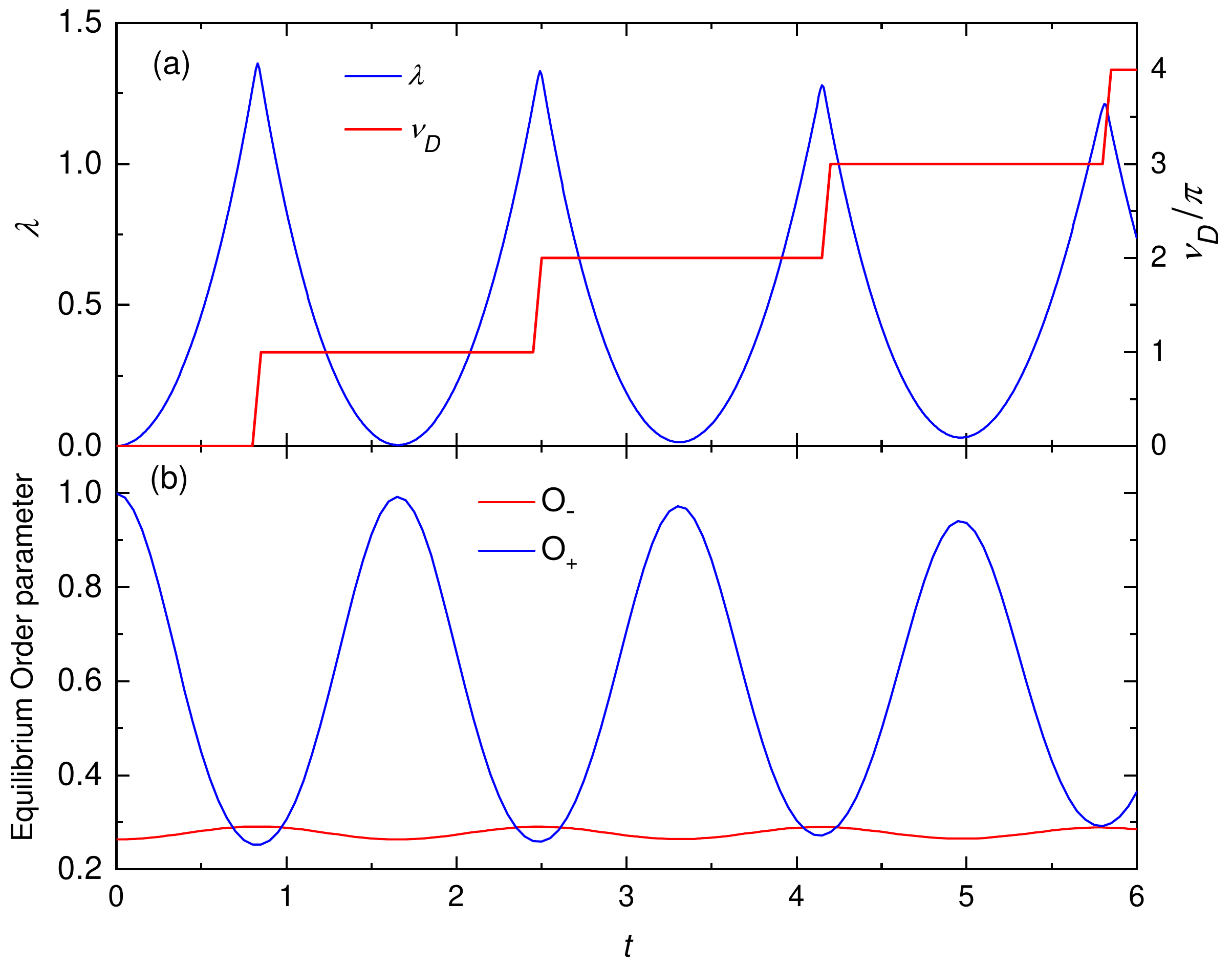}
	\caption{The time dependence of (a) the Loschmidt rate (left y-axis) and the DTOP (right y-axis), and (b) the equilibrium order parameters $O_{\pm}$ in the non-interacting SSH model for quench from $\eta_0=0.9$ to $\eta_1=-0.9$. Periodic boundary condition is used and $N=100$.}
	\label{fig1a}
\end{figure}

\begin{figure}
	\includegraphics[width=0.5\textwidth]{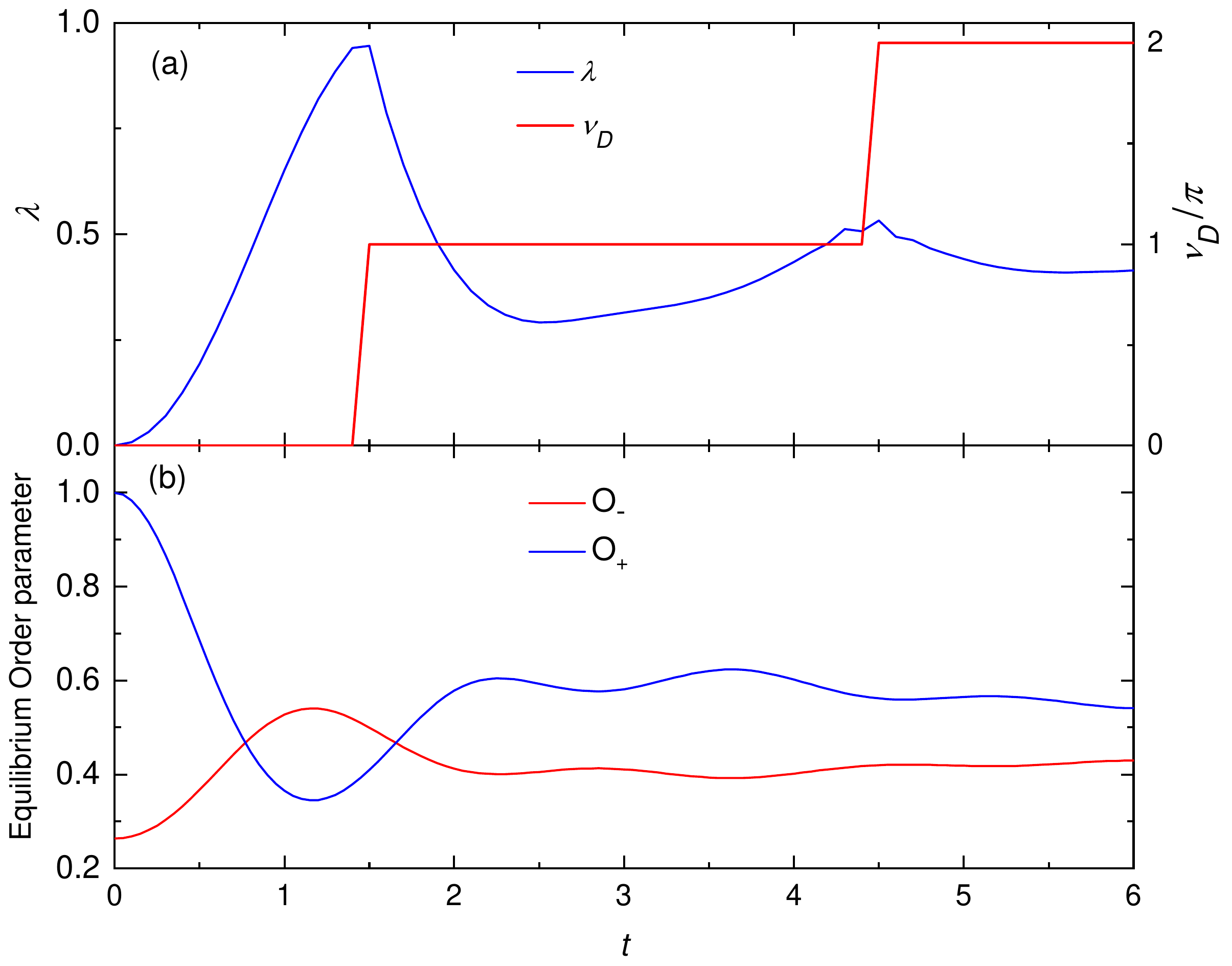}
	\caption{The time dependence of (a) the Loschmidt rate (left y-axis) and the DTOP (right y-axis), and (b) the equilibrium order parameters $O_{\pm}$ in the non-interacting SSH model for quench from $\eta_0=0.9$ to $\eta_1=-0.3$. Periodic boundary condition is used and $N=100$}
	\label{fig1b}
\end{figure}



\section{Quenched non-interacting SSH model}
\label{sec:non-interacting}

\begin{figure}
	\centering
	\includegraphics[width=0.5\textwidth]{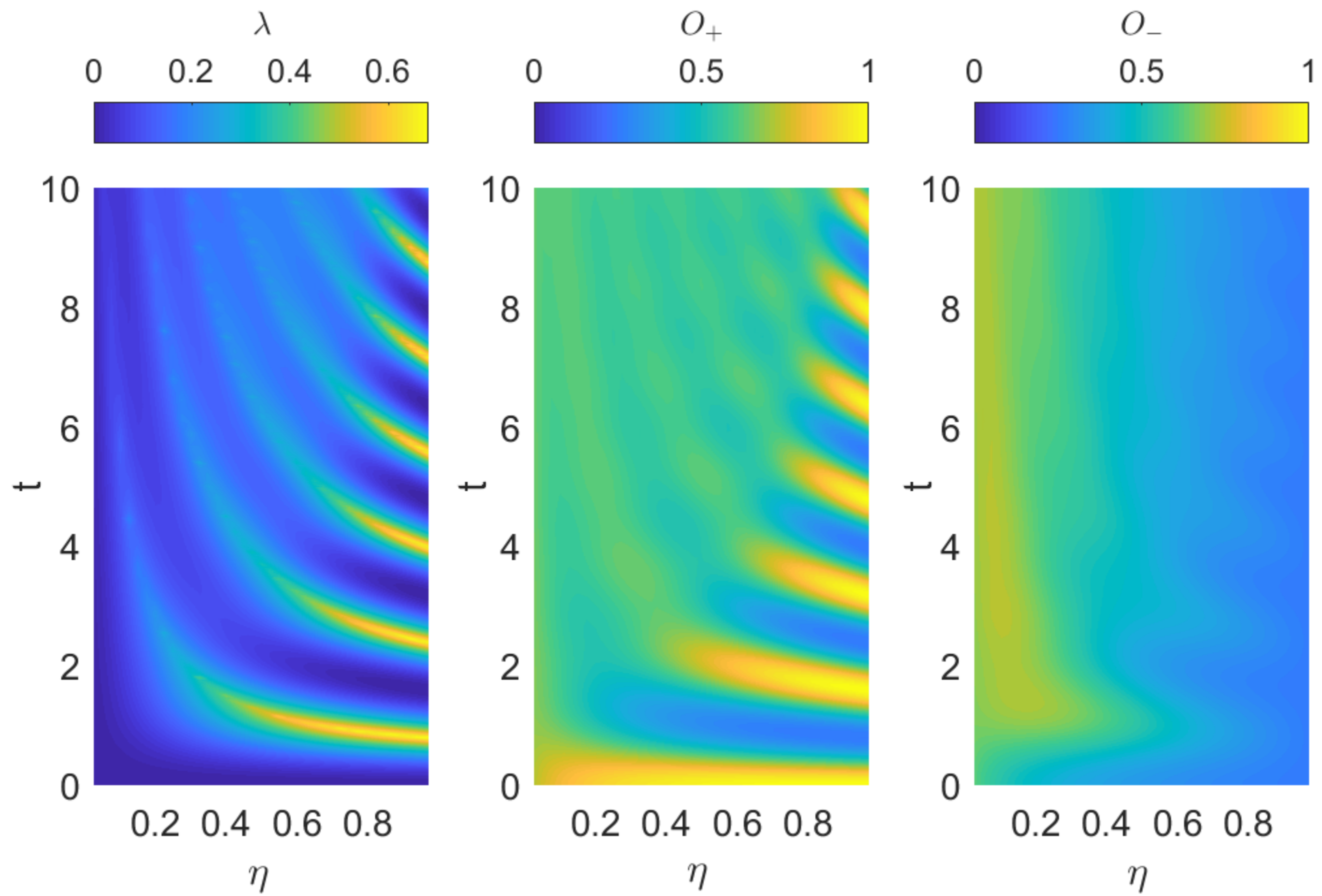}
	\caption{Colormap of the Loschmidt rate (left panel), $O_+$ (middle panel), and $O_-$ (right panel) for a quench from the trivial phase to the topological phase. Here $N=100$.}
	\label{fig:map_noninteract}
\end{figure}

\begin{figure}
	\includegraphics[width=0.4\textwidth]{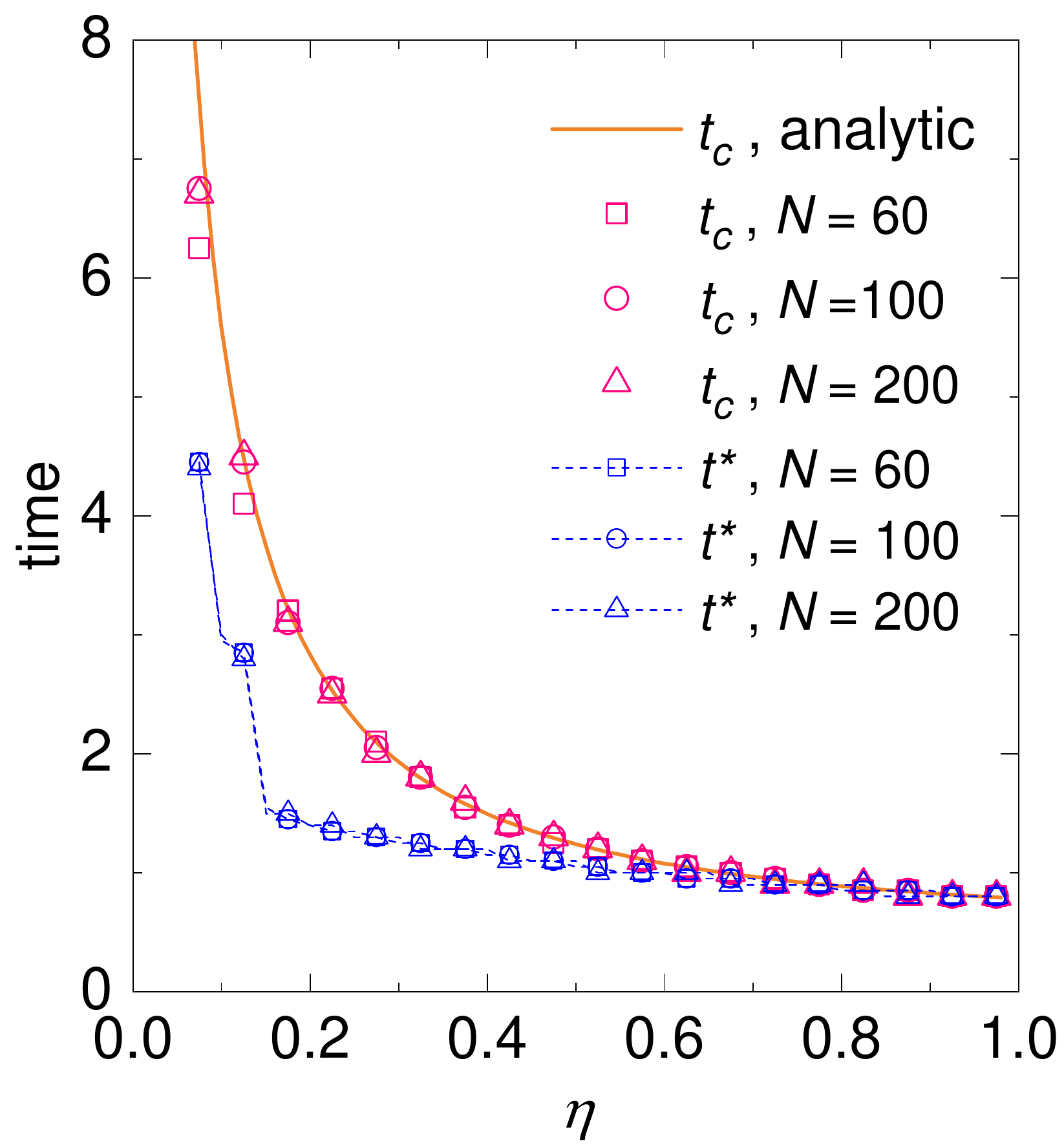}
	\caption{The critical time $t_c$, as detected by the first peak in the rate function, and the first minimum of $O_+$ occurrence time $t^*$, as a function of $|\eta_0|=|\eta_1|=\eta$ for a symmetric quench for various system sizes. The solid orange line shows the critical time in the thermodynamic limit. The dashed blue lines are just guides to the eyes. }
	\label{fig:tc_eta}
\end{figure}

In the following, unless otherwise specified, we consider the system prepared in an initial state given by the ground state $\ket{\Psi(0)}$ of the Hamiltonian at $g_0 = (\eta_0,U_0,V_0)$. At $t=0$, the system's parameters are quenched to $g_1 = (\eta_1,U_1,V_1)$, and the time evolution of the state follows $\ket{\Psi(t)}=e^{-iH(g_1)t}\ket{\Psi_0}$.

First, we considered the case without interaction, i.e. $U=V=0$. The initial state is prepared in the ground state of the trivial phase and the system is quenched across the underlying equilibrium topological quantum phase transition point at $\eta=0$. Figure \ref{fig1a} shows the Loschmidt rate and the order parameters in Eq. (\ref{eq:O-}) and (\ref{eq:O+}) as a function of time for a quench $\eta_0=0.9\rightarrow\eta_1=-0.9$ respectively. Periodic boundary condition is used. The result for the Loschmidt rate is consistent with Ref. \cite{Sedlmayr2018}. For this symmetric quench from a state deep inside the trivial phase, the critical times $t_c$ are evenly separated as indicated by the periodic peaks in the rate function. Since we started from an initial state in the trivial phase, the order parameter $O_+$ is maximum at $t=0$.  It shows periodic oscillation in the transient time and a minimum occurs at the DQPT. At these critical times, $O_-$ becomes maximum, suggesting that the state has the strongest topological character.

Figure \ref{fig1a}(a) also shows the DTOP defined in terms of the Pancharatnam geometrical phase (PGP) \cite{Budich2016}. For non-interacting models, the Loschmidt amplitude can be decomposed into $l(t)=\prod_{k>0}l_k(t)$. The PGP can be calculated by $\phi_k^G(t)=\phi_k(t)-\phi_k^{d}(t)$, where $\phi_k(t)$ is the polar angle of $l_k(t)$, and $\phi_k^d(t)=-\int_0^tds\mele{\psi_k(s)}{H(g_1)}{\psi_k(s)}$ is the dynamical phase. The DTOP is then defined as
\bea
\nu_D(t)=\frac{1}{2\pi}\oint_0^{\pi}\frac{\partial\phi_k^G(t)}{\partial k}.
\eea 
As seen from the figure, $\nu_D$ attains quantized values and it jumps by $\pi$ at the critical time $t_c$ as detected by the Loschmidt rate peaks.

We also considered the case of an asymmetric quench where $|\eta_0|\ne|\eta_1|$ as shown in Fig \ref{fig1b}. The initial state is prepared at $\eta_0=0.9$, which is deep inside the trivial phase in equilibrium, and quenched to $\eta_1=-0.3$. Unlike the symmetric quench case, the Loschmidt rate and the time evolution of the order parameters become aperiodic. While the first local extrema of $O_{\pm}$ leads the first Loschmidt rate peak, the time at which the DTOP jumps still agrees with the Loschmidt rate peak.


Figure \ref{fig:map_noninteract} shows the colormaps of the Loschmidt rate and the equilibrium order parameters for a quench from $\eta_0=\eta>0$ to $\eta_1=-\eta_0 $. The trend of $t_c$, as indicated by the Loschmidt rate peak, agrees qualitatively with the $t^*$, as indicated by the minima of the $O_+$. A quantitative comparison of $t_c$ and $t^*$ is presented in Fig. \ref{fig:tc_eta}. Both the $t_c$ and $t^*$ extracted from numerical calculation do not change significantly with the system size. Moreover, the extracted $t_c$ agrees well with the one in the thermodynamic limit, which can be obtained analytically with the expression 
\bea
t_c = \frac{\pi\sqrt{1+\eta^2}}{4\sqrt{2}\eta}.
\eea
For large $|\eta|$, $t_c$ and $t^*$ agrees well but they start to deviate around  $|\eta|=0.6$. Moreover, from the rightmost panel of Fig. \ref{fig:map_noninteract}, we can see that $O_-$ in general grows up at $|\eta|$ around zero. From the expression of $O_-$, we can deduce that inter-cell correlations start to build up around this region, and thus giving rise to a stronger entanglement in the system.

\begin{figure}
	\centering
	\includegraphics[width=0.5\textwidth]{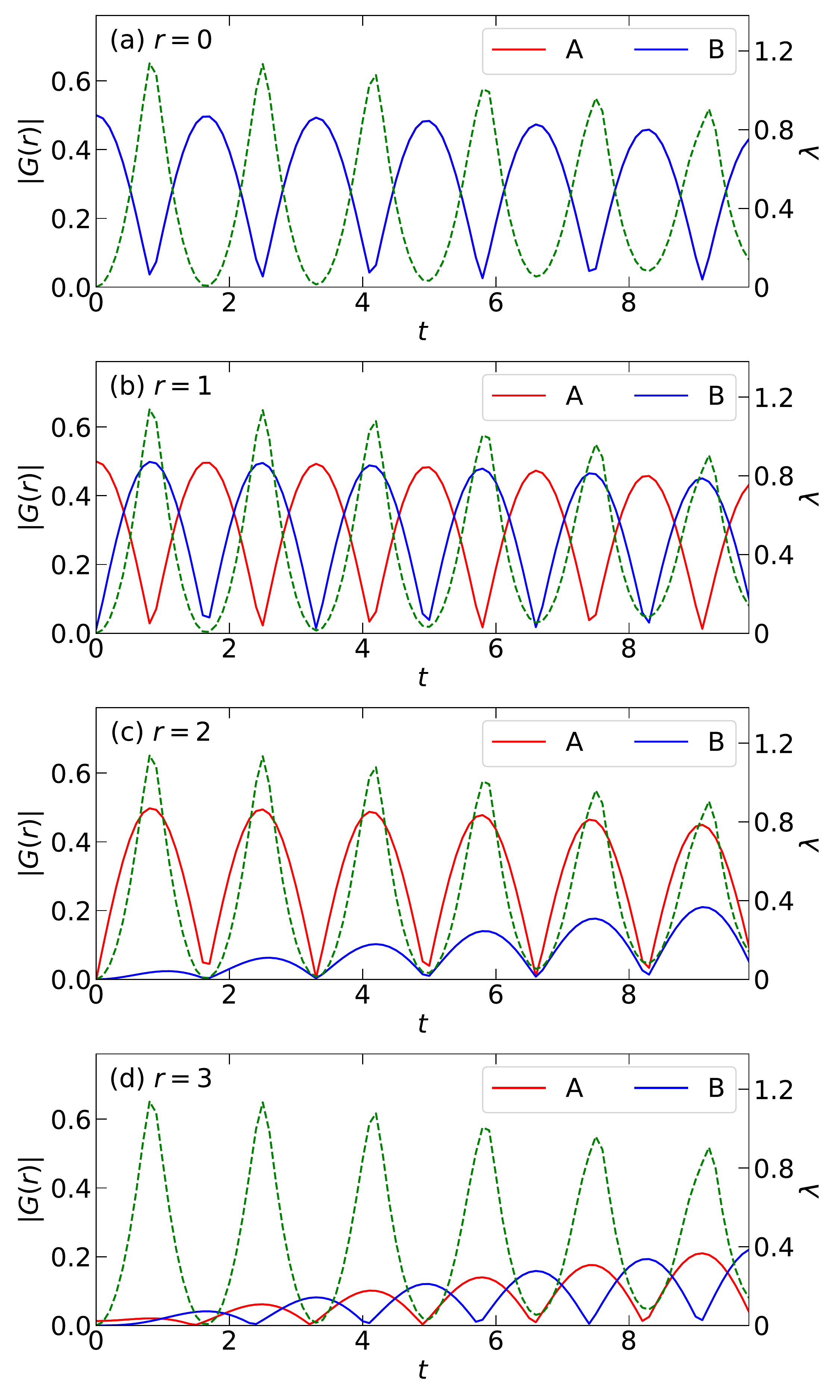}
	\caption{Solid lines show the magnitude of the unequal time two-point Green function in the non-interacting SSH model for a quench from $\eta=0.9$ to $\eta=-0.9$. Dash green lines show the Loschmidt rate $\lambda$ as a function of time. Here $N=10$.}
	\label{fig:GF_symm}
\end{figure}

\begin{figure}
	\centering
	\includegraphics[width=0.5\textwidth]{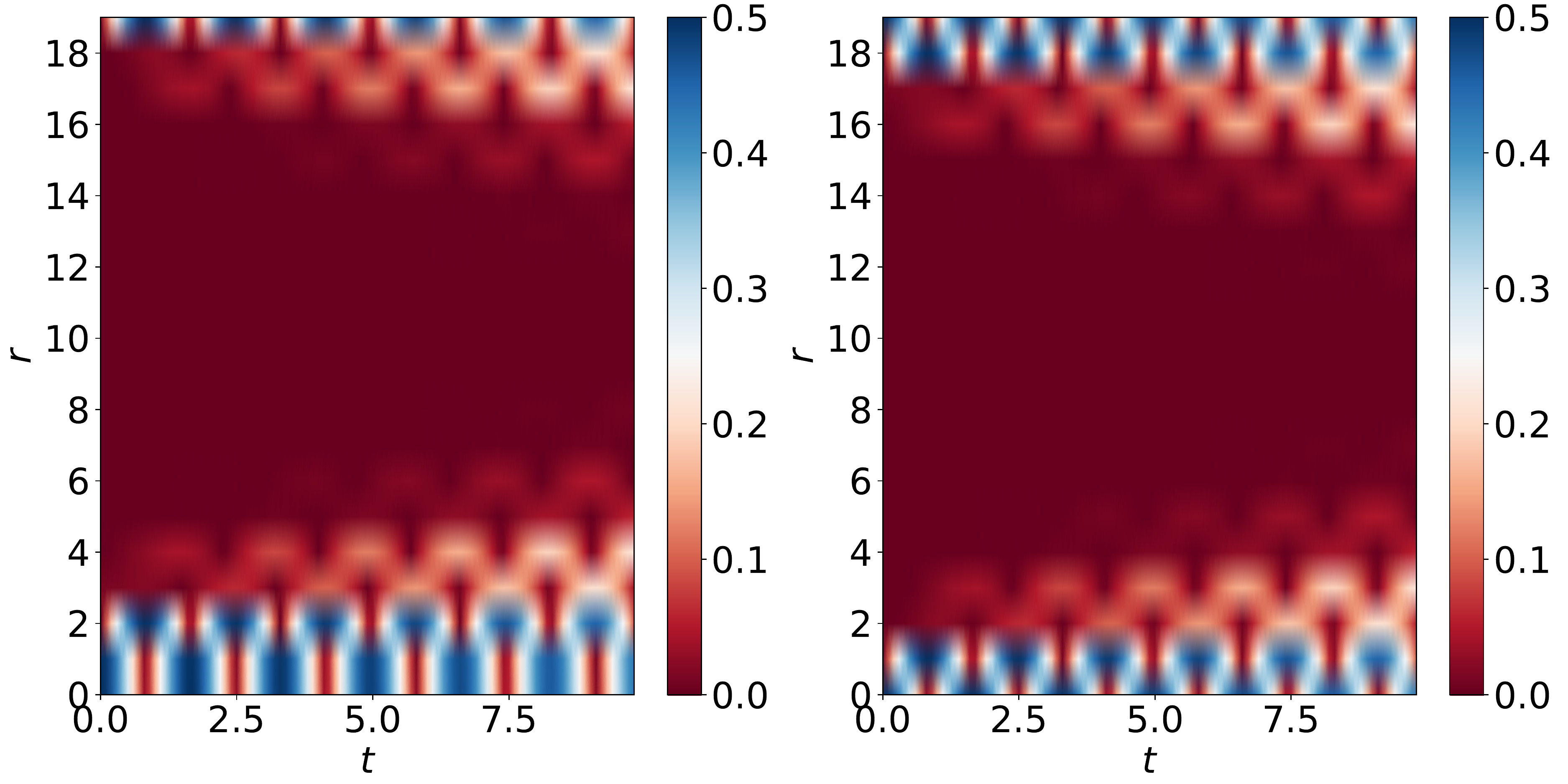}
	\caption{Colormap of the magnitude of the unequal time two-point correlation $|G_A(r,t)|$ (left panel) and $|G_B(r,t)|$ (right panel) as a function of time and separation for the non-interacting SSH model. Here the quench is from $\eta_0=0.9$ to $\eta_1=-0.9$ and $N=10$. }
	\label{fig:map_GF_symm}
\end{figure}

\begin{figure}
	\centering
	\includegraphics[width=0.5\textwidth]{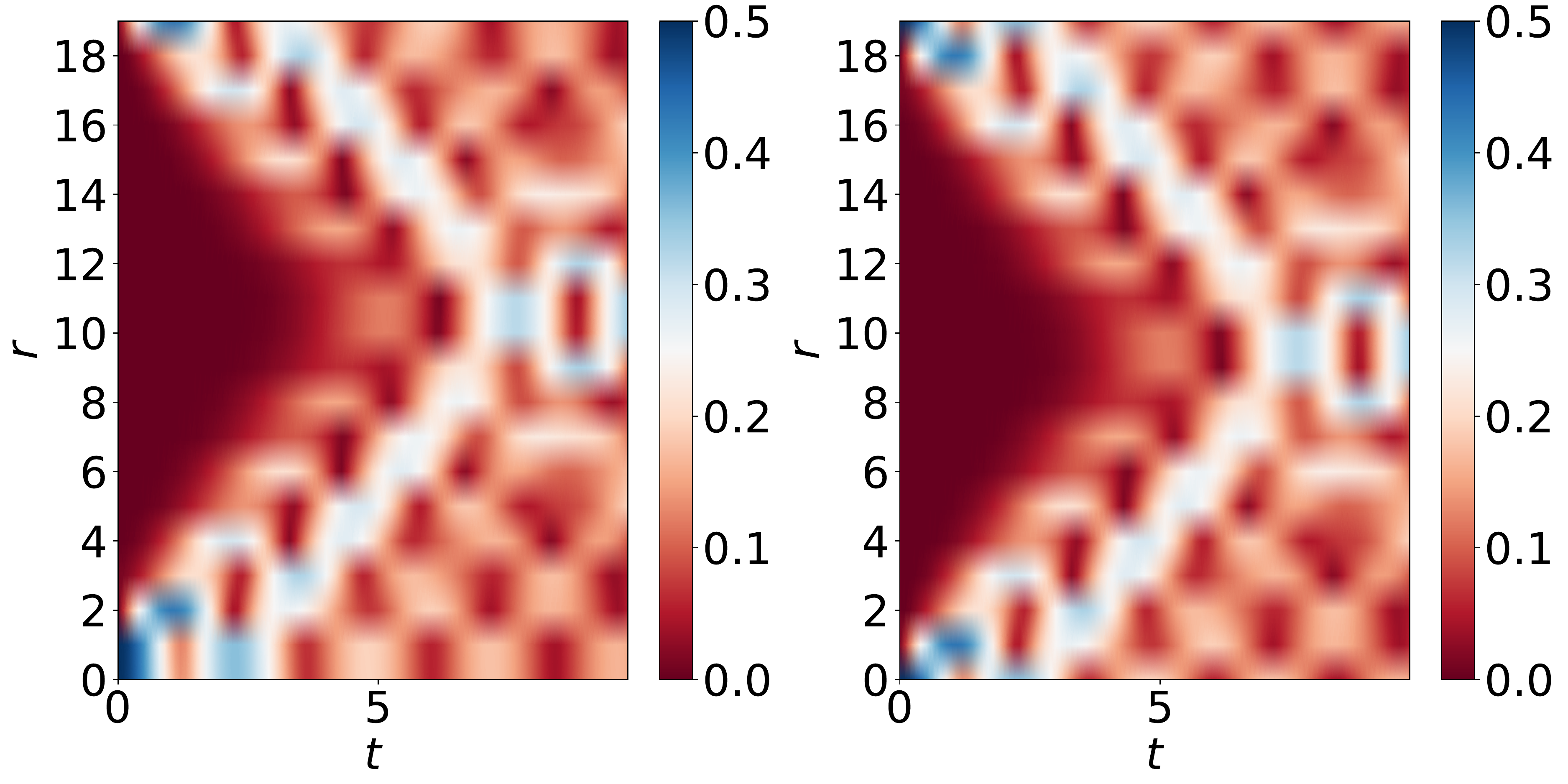}
	\caption{Colormap of the magnitude of the unequal time two-point correlation $|G_A(r,t)|$ (left panel) and $|G_B(r,t)|$ (right panel) as a function of time and separation for the non-interacting SSH model. Here the quench is from $\eta_0=0.9$ to $\eta_1=-0.3$ and $N=10$.}
	\label{fig:map_GF_asymm}
\end{figure}

To further investigate how the correlation builds up in the system, we calculated the Green function, i.e. the unequal time two-point correlation, as defined by
\bea
G_{A/B}(r,t)&=&\mele{\Psi(0)}{c_r(t)c^{\dagger}_{0,A/B}}{\Psi(0)},\\ \nonumber
&=&\mele{\Psi(0)}{e^{iH(g_1)t}c_re^{-iH(g_1)t}c^{\dagger}_{0,A/B}}{\Psi(0)}
\eea
where $r$ measures the distance from the A/B site in the first unit cell and the separation between adjacent sublattices is taken to be one.

Figure \ref{fig:GF_symm} shows the magnitude of the Green function for the first few nearest neighbours in the symmetric quench from $\eta_0=0.9$ to $\eta_1=-0.9$. Physically, the Green function measures the overlap between the two states $e^{-iH(g_1)t}c_{0,A/B}^{\dagger}\ket{\Psi(0)}$ and $c_r^{\dagger}e^{-iH(g_1)t}\ket{\Psi(0)}$. The former state is the case in which we create a fermion at site A/B in the first unit cell and allow the system to evolve for a time $t$ with the quench Hamiltonian. The latter state is the case in which we first allow the system to evolve for a time $t$ and create a fermion at a site with distance $r$. When $r=0$, the Green function basically tells us how likely the system can go back to the initial state if the creation and evolution process  is reversed. It should thus resemble the feature of the Loschmidt echoes. Therefore, the minimum of $G_{A/B}(r=0,t)$ shall occur at the Loschmidt rate peak as expected.  Moreover, at the DQPT, the correlation between the first A site and first B site as measured by $G_A(r=1,t)$ is minimum while that for the first A site and the second A site is maximum as indicated by $G_A(r=2,t)$. This suggests that the quench here is strong enough to turn the system into the topological state at the transition.

From Fig. \ref{fig:GF_symm} and Fig.\ref{fig:map_GF_symm}, we also observe that for $r>0$, the correlation is mainly dominated by the nearest and the next nearest neighbors in the beginning. However, as time increases, the correlation with further neighbours also grows. Similar conclusion was also reached by considering the equal-time two-point correlation \cite{Sedlmayr2018}.  The behavior of the Green function at $r=N/2+m$, where $m=1,2,\cdots$ is the same as that at $r=N/2-m$ with the role of $|G_A|$ and $|G_B|$ exchanged. This is just a consequence of using periodic boundary condition. We also considered a symmetric quench from the trivial phase to the topological phase, i.e. from $\eta_0=-0.9$ to $\eta_1=0.9$, the same results are obtained but just with the role of $|G_A|$ and $|G_B|$ interchanged. On the contrary, if the system is quenched from the trivial phase $\eta_0=0.9$ to a shallower (closer to the underlying equilibrium transition point) topological phase, the correlation with the nearest neighbor decays faster while the correlation with the neighbors further away grows faster as compared to the quench with $\eta_1$ deep inside the phase. This is illustrated in Fig.\ref{fig:map_GF_asymm}, in which a quench from $\eta_0=0.9$ to $\eta_1=-0.3$ is considered.


\section{Interaction effect on the quench dynamics}\label{sec:interacting}

\begin{figure}
	\centering
	\includegraphics[width=0.5\textwidth]{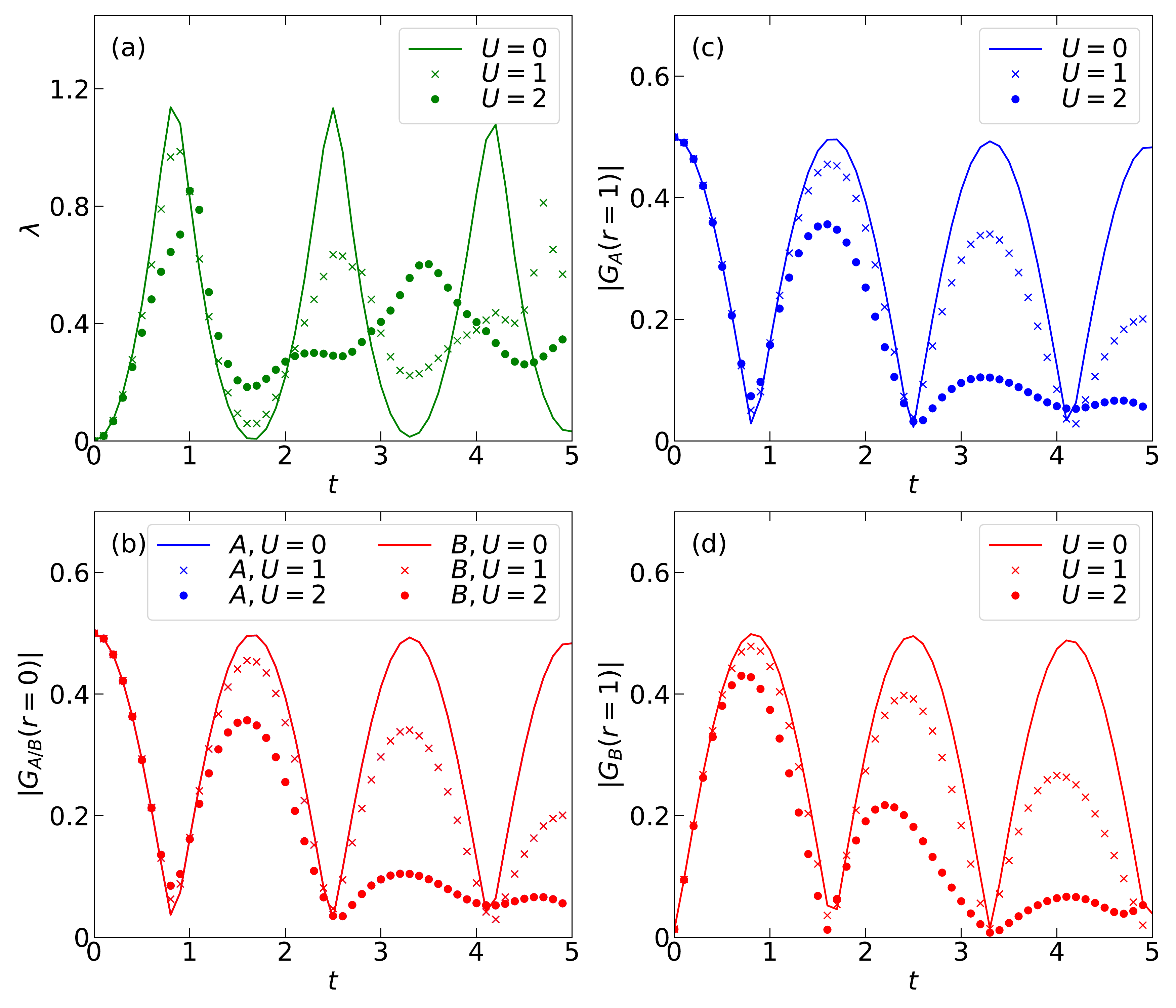}
	\caption{(a) The Loschmidt rate and (b)-(d) the magnitude of the Green function as a function of time in the case of quenching the SSH model from $g_0=(0.9,0,0)$ to $g_1=(-0.9,U,0)$. Here $N=10$.}
	\label{fig:int_U}
\end{figure}

\begin{figure}
	\centering
	\includegraphics[width=0.5\textwidth]{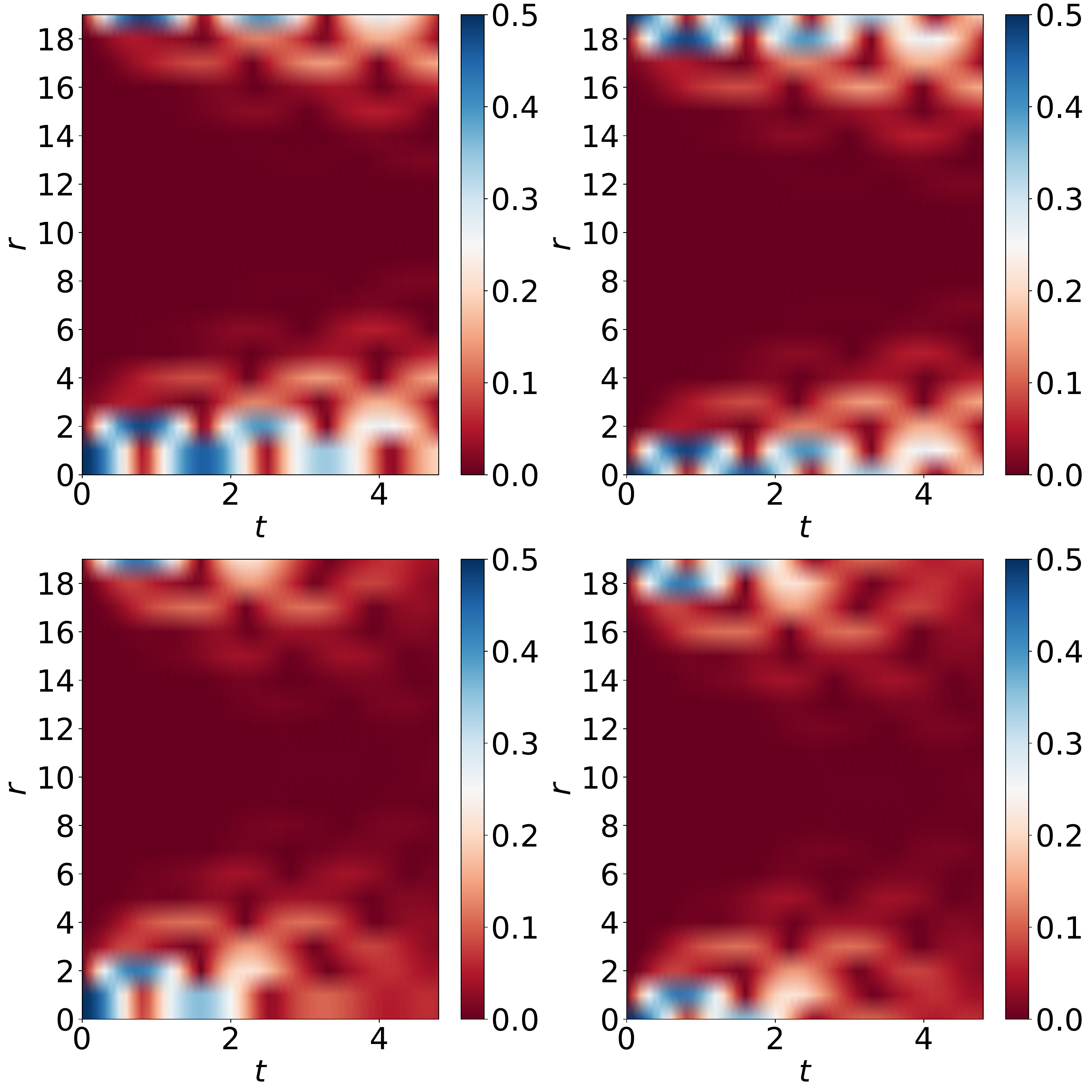}
	\caption{Colormap of the magnitude of the Green function as a function of time and separation for the SSH model with interactions. The initial state is the ground state of the Hamiltonian at $g_0=(0.9,0,0)$. Top panel show the quench to $g_1=(-0.9,1,0)$ and the bottom panel shows the quench to $g_1=(-0.9,2,0)$. Left panel and the right panel corresponds to $|G_A(r,t)|$ and $|G_B(r,t)|$, respectively.}
	\label{fig:map_int_U}
\end{figure}

In this section, we consider the effect of interaction on the system around the DQPT. Figure \ref{fig:int_U} and \ref{fig:map_int_U} show the Green functions for the case of quenching the SSH model from $g_0=(0.9,0,0)$ to $g_1=(-0.9,U,0)$. From Fig. \ref{fig:int_U}, we note that upon the addition of repulsive intra-cell interaction $U$, the first critical time as signalled by the Loschmidt rate peak increases slightly as $U$ increases. On the contrary, the time at which the first minimum in $|G_{A/B}(r=0)|$ and $|G_{A}(r=1)|$ does not seem to affect much by $U$, while $|G_B(r=1)|$ shows its first maximum at a slightly earlier time as $U$ increases. This suggests that in this short time regime around the first DQPT, while the intra-cell interaction does not have much effect on suppressing the intra-cell correlation, it accelerates the build up of the nearest inter-cell correlation. Interestingly, this in turn delays the occurrence of the first DQPT. As time progresses, we can see that the interaction generally suppresses the nearest neighbour correlation away from the DQPT. The decay of the nearest neighbour correlation is faster when the interaction $U$ is stronger.  On the other hand, by comparing Fig. \ref{fig:map_int_U} with Fig. \ref{fig:map_GF_symm}, we can see that the correlations with sites that are further away build up more quickly with the addition of $U$.

\begin{figure}
	\centering
	\includegraphics[width=0.5\textwidth]{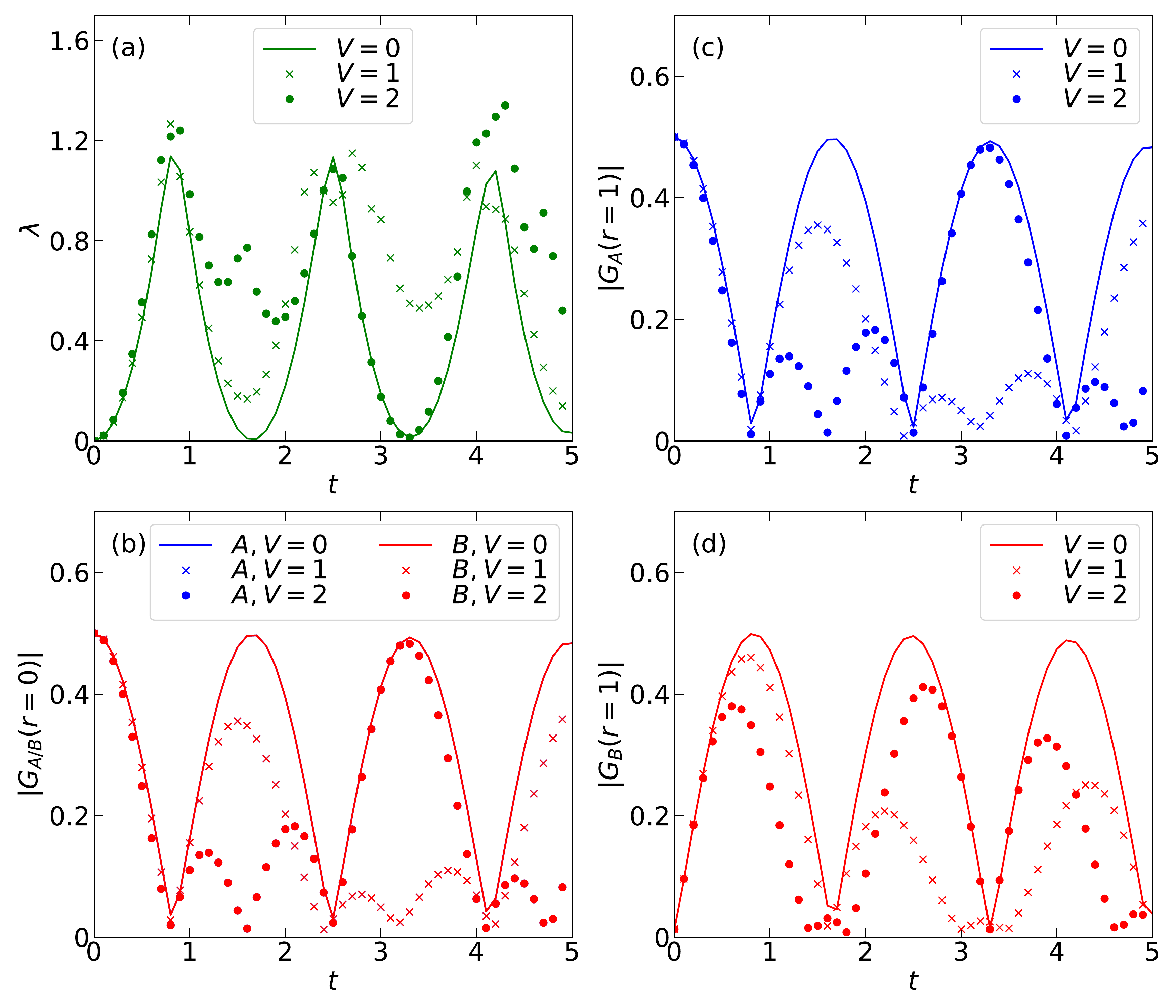}
	\caption{(a) The Loschmidt rate and (b)-(d) the magnitude of the Green function as a function of time in the case of quenching the SSH model from $g_0=(0.9,0,0)$ to $g_1=(-0.9,0,V)$. Here $N=10$.}
	\label{fig:int_V}
\end{figure}

\begin{figure}
	\centering
	\includegraphics[width=0.5\textwidth]{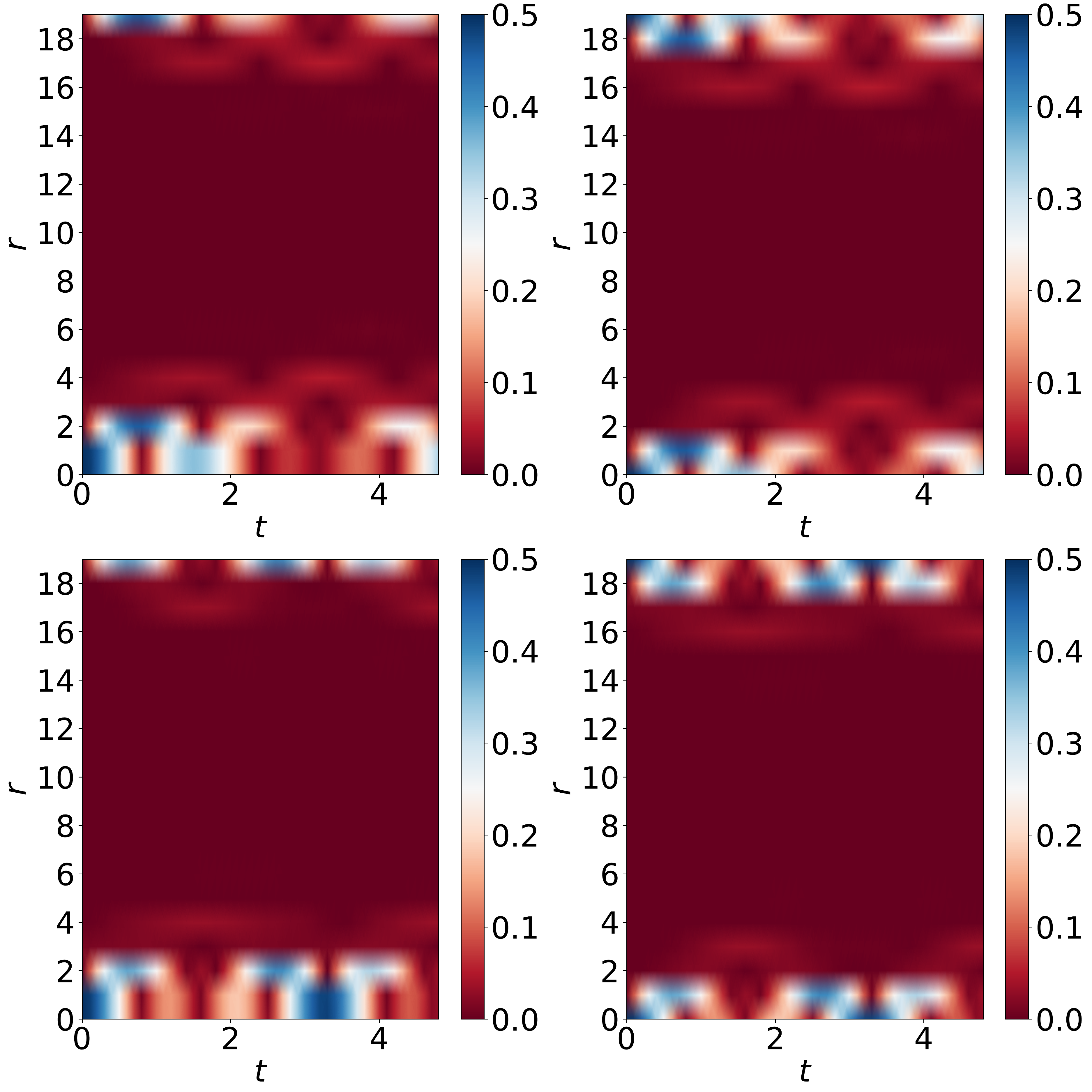}
	\caption{Colormap of the magnitude of the Green function as a function of time and separation for the SSH model with interactions. The initial state is the ground state of the Hamiltonian at $g_0=(0.9,0,0)$. Top panel show the quench to $g_1=(-0.9,0,1)$ and the bottom panel shows the quench to $g_1=(-0.9,0,2)$. Left panel and the right panel corresponds to $|G_A(r,t)|$ and $|G_B(r,t)|$, respectively.}
	\label{fig:map_int_V}
\end{figure}

The effect of adding inter-cell interactions $V$ is also considered. Figure \ref{fig:int_V} and \ref{fig:map_int_V} show the dynamics in the model for a quench from $g_0=(0.9,0,0)$ to $g_1=(-0.9,0,V)$. We can see that the first critical time and the first minimum of the intra-cell correlation are insensitive to the addition of repulsive inter-cell interactions. However, the nearest inter-cell correlation $|G_B(r=1)|$ grows to its first maximum faster and occurs before the DQPT as detected by the Loschmidt rate when $V$ increases. Different from the case of adding the intra-cell interaction $U$ in which the extrema of the nearest neighbours ($r=1$) takes place at a similar time, the Green function oscillates with a more irregular pattern here. As time grows, Fig. \ref{fig:map_int_V} shows that the correlation between any two sites are suppressed in general as compared to the non-interacting case, and no obvious blow up of the long-range two-point correlation is observed in contrast to the case in the presence of intra-site interaction $U$.

\section{Quench across equilibrium phase boundary in the interacting case}
\label{sec:interacting2}
 
\begin{figure}
	\includegraphics[width=0.48\textwidth]{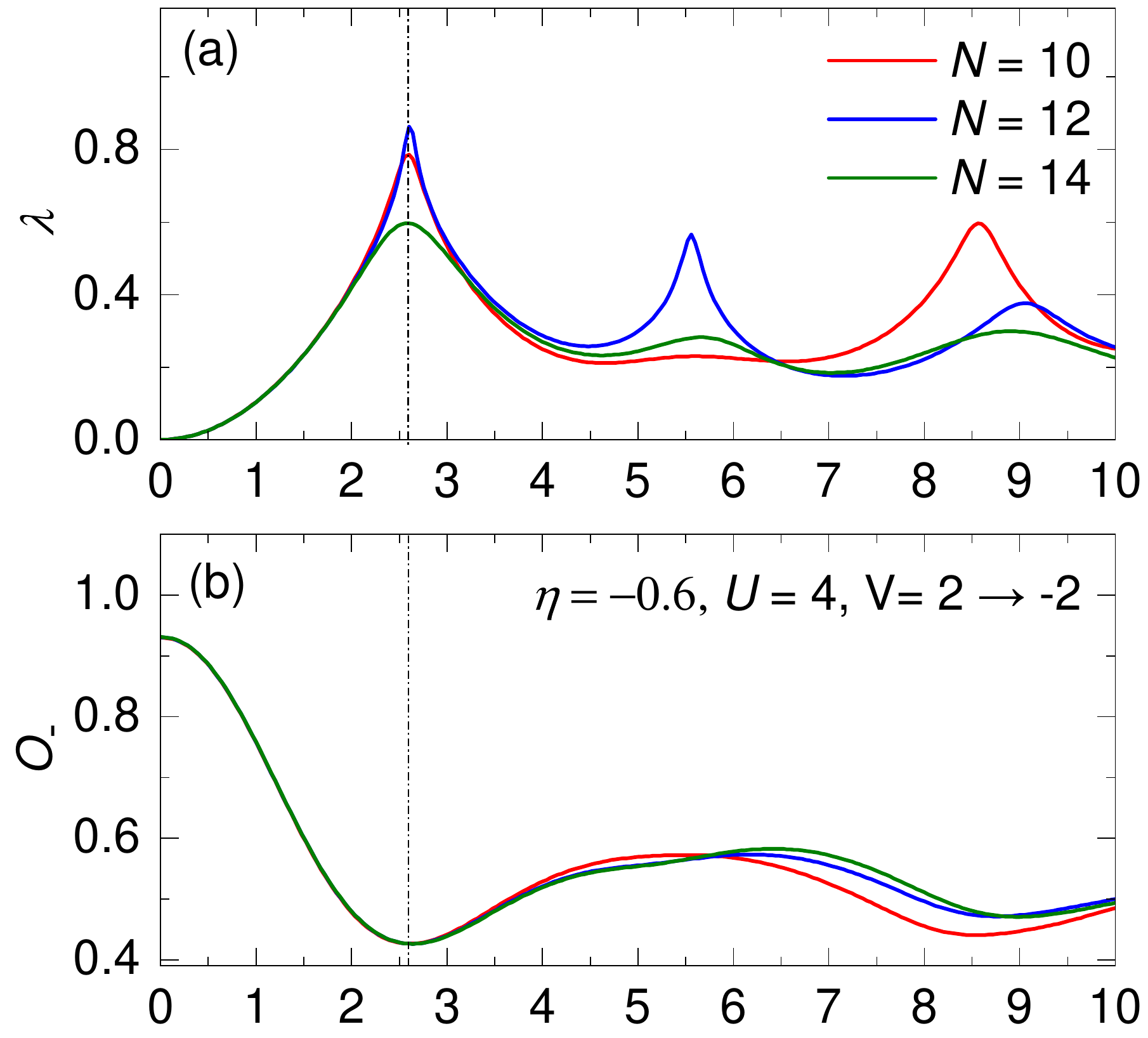}
	\caption{The time dependence of (a) the Loschmidt rate and (b) the equilibrium order parameter $O_-$  in the interacting SSH model for quench from the topological phase to the trivial phase for various system size. Here $g_0=(-0.6,4,2)$, $g_1=(-0.6,4,-2)$.}
	\label{fig:topo_trivial}
\end{figure}

\begin{figure}
	\includegraphics[width=0.48\textwidth]{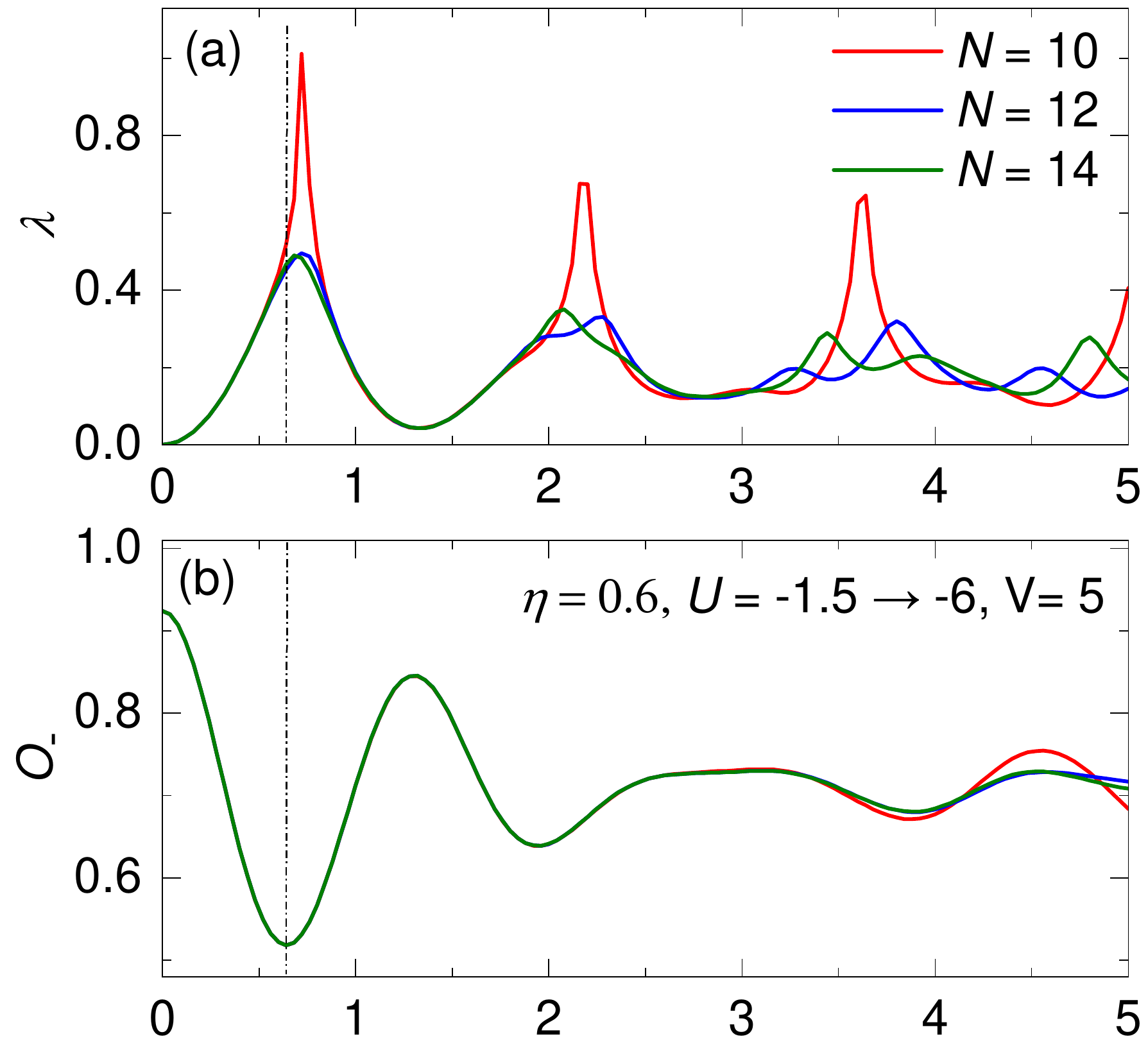}
	\caption{The time dependence of (a) the Loschmidt rate and (b) the equilibrium order parameter $O_-$ in the interacting SSH model for quench from the topological phase to the CDW phase for various system sizes. Here $g_0=(0.6,-1.5,5)$, $g_1=(0.6,-6,5)$.}
	\label{fig:topo_CDW}
\end{figure}

\begin{figure}
	\includegraphics[width=0.48\textwidth]{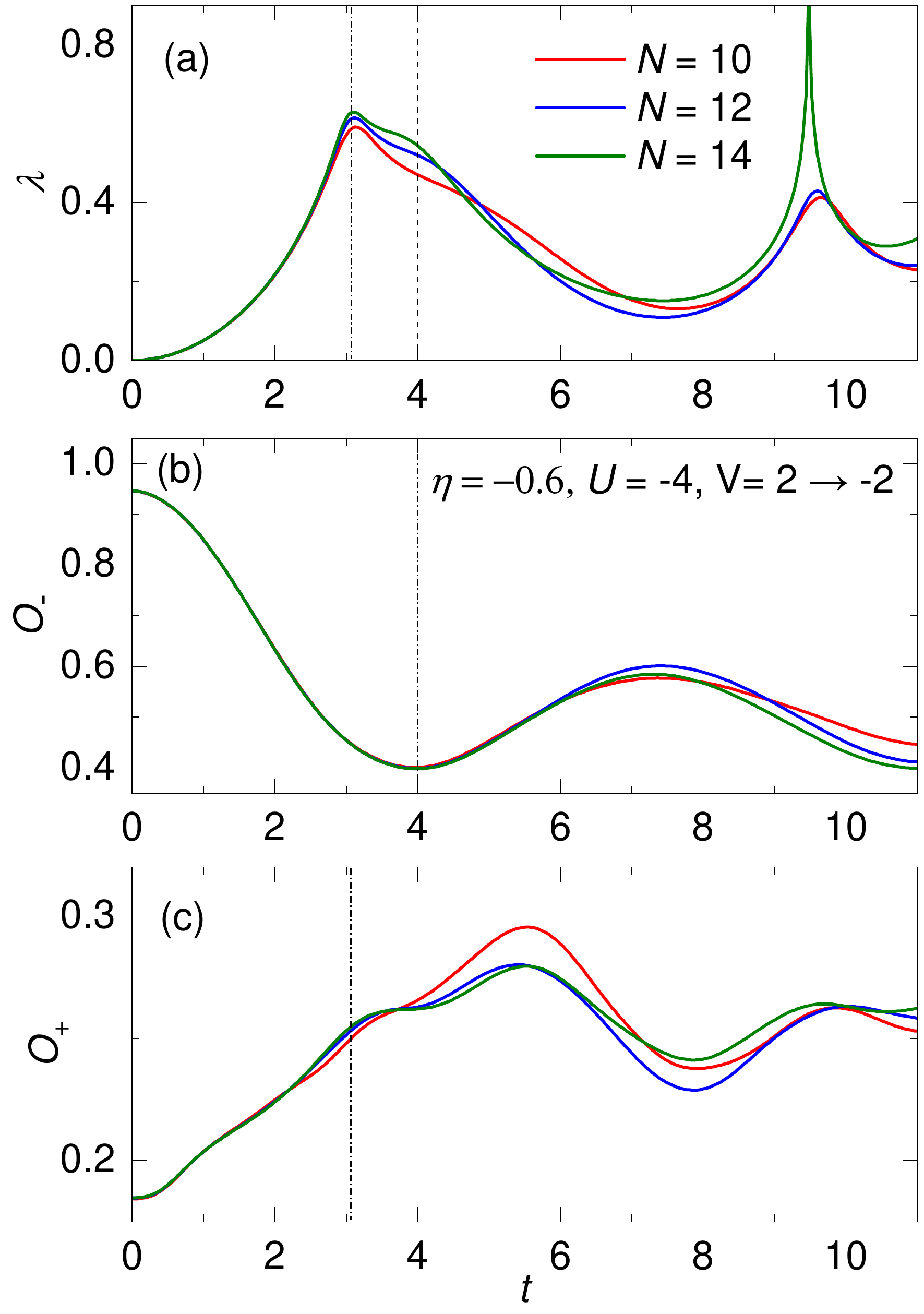}
	\caption{The time dependence of (a) the Loschmidt rate, (b) the equilibrium order parameter $O_-$, and (c) $O_+$ in the interacting SSH model for quench from the topological phase to the PS phase for various system sizes. Here $g_0=(-0.6,-4,4)$ and $g_1=(-0.6,-4,-4)$.}
	\label{fig:topo_PS} 
\end{figure}

Next, we consider the quench across various phases in the equilibrium ground-state phase diagram of the interacting SSH model. The corresponding phase diagram can be found in Ref. \cite{Yu2016}. Three cases are considered as listed below. 

\textit{\textbf{Quench from topological phase to trivial phase -}} Figure \ref{fig:topo_trivial} shows the results for quench from $(\eta_0, U_0, V_0)=(-0.6,4,2)$ to $(\eta_1, U_1, V_1)=(-0.6,4,-2)$. As compared to the non-interacting case, the rate function becomes more irregular. Nevertheless, the first transition time as indicated by the Loschmidt rate peak is insensitive to the system size whereas the finite size effect starts to kick in only after the first transition. The order parameter $O_-$ decreases from an initial value close to one as expected since we start with an initial state in the topological phase. At the first critical time as indicated by the Loschmidt rate peak, $O_-$ becomes minimum, and this minimum does not change significantly with the system size. Obviously, the coincidence in $t_c$ and $t^*$ will remain in a large system. This suggests that the topological order in the system is suppressed to the greatest extent exactly when the first DQPT takes place.

\textit{\textbf{Quench from topological phase to CDW phase -}} Figure \ref{fig:topo_CDW} shows the results for quench from $(\eta_0, U_0, V_0)=(0.6,-1.5,5)$ to $(\eta_1, U_1, V_1)=(0.6,-6,5)$.  Similar to the case above in which the system is quenched from the topological phase to the trivial phase, the first non-analyticity in the Loschmidt rate takes place around the minimum of the $O_-$ order parameter. As the system size increases, the Loschmidt rate peak occurs at a slightly earlier time, and agrees better with the time in which $O_-$ shows a minimum. We can expect that at the first DQPT, the topological order in the system again attains a minimum.

\textit{\textbf{Quench from topological phase to PS phase - }} Figure \ref{fig:topo_PS} shows the results for quench from $(\eta_0, U_0, V_0)=(-0.6,-4,4)$ to $(\eta_1, U_1, V_1)=(-0.6,-4,-4)$. In contrast to the previous cases, a second non-analyticity appears in the Loschmidt rate around its first peak as the system size becomes large. This suggests two DQPTs ocurrings at a close vicinity. Interestingly, although the time at which the Loschmidt rate's first singularity peak occurs is different from that of the first minimum of $O_-$, its second singularity tends to align with the minimum of $O_-$ when the system size increases (as indicated by the dash vertical line in the figure). In addition, the first singularity peak of Loschmidt rate correlates with the change of the $O_+$ order. One can see that there is a round weak peak near $t\approx3$ (as indicated by the dash line in (c)), and this peak generally aligns with the first sharp peak in the Loschmidt rate.

\section{Conclusions}
\label{sec:conclusion}

We studied the dynamical quantum phase transitions in the SSH model with various measures of the correlations in the system. Different quench parameters are considered. 

In the non-interacting case, we find that the equilibrium order parameter $O_{\pm}$ shows an extrema around the DQPT. The alignment between the Loschmidt rate peaks and the equilibrium order parameter agree well when the quench is to and from a state deep inside a phase. However, slight deviation starts to occur when the quench is in a shallow phase and in particular, the order parameter extremum is found to lead the transition time. We attribute this to the presence of non-local correlation in the system. This may be intrinsically connected to the two notions of precession and entanglement DQPTs currently introduced by S. De Nicola et. al. and further exploration along this direction will be of interest \cite{DeNicola2021}. We also investigated the unequal time two-point correlations in the system. In the short-time regime, the nearest and next nearest neighbor correlations dominates and as time increases, the correlation with neighbors further away builds up. The growth of further site correlation is found to be faster in the asymmetric quench case where the system is quenched across the phase boundary to a shallow phase.

We also considered the quenched SSH model with interactions. Upon the addition of repulsive intra-cell interaction, the first DQPT was found to be delayed. In contrast to the non-interacting case, the unequal time nearest neighbor correlation decays faster while the further neighbor correlations builds up more quickly.  On the other hand, the effect on the first DQPT in the presence of inter-cell interaction seem to be insignificant. In general, the addition of inter-cell interaction suppresses the two-point correlation and there is no obvious build up of the correlations between sites further away even as time grows.

The quench dynamics across the underlying equilibrium quantum phase transition in the interacting SSH model is also investigated. The equilibrium order parameter in general present signatures around the DQPT. While the equilibrium order parameter $O_-$ shows a minimum exactly at the first DQPT when quenching the system from the topological phase to the trivial phase and to the CDW phase in the presence of interactions, the first minimum of $O_-$ lags behind the Loschmidt rate's first peak for the quench from topological phase to the phase separation phase. 
The time at which the first minimum of $O_-$ occurs instead aligns with the second non-analyticity in the Loschmidt rate. It will be worthy to investigate the physical reasons behind these observations in future works. 
Lastly, we shall mention that despite the simulation was carried out with an initial state in the topological phase in the interacting case, we shall expect that similar observations will result if we start with a state in the trivial phase but the role of $O_+$ and $O_-$ will be exchanged if the parameter values are replaced from $(\eta, U, V)$ to $(-\eta,V,U)$.

\section{Acknowledgement}
We acknowledge financial support from Research Grants Council of Hong Kong (Grant No. ECS/21304020), City University of Hong Kong (Grant No. 9610438), FCT Grant UID/CTM/04540/2019, National Natural Science Foundation of China (Grant No. 12074376 and 12088101), NSAF (Grant No. U1930402), and computational resources from the Beijing Computational Science Research Center.

\end{document}